\pdfoutput=1
\documentclass[letterpaper,journal]{IEEEtran}
\usepackage{amsmath,amssymb,amsfonts}
\usepackage{array}
\usepackage[caption=false,font=footnotesize]{subfig}
\usepackage{textcomp}
\usepackage{url}
\usepackage{graphicx}
\usepackage{cite}
\usepackage{tabularx}
\usepackage{placeins}
\begin{document}
\title{Soft Temporal Alignment Distillation for Camera-to-mmWave Radar Human Activity Recognition}

\author{Jiuqi Yan, Chengbo Zhu, Chengdong Xu and Qisong Wu, \IEEEmembership{Senior Member, IEEE}%
\thanks{This work has been submitted to the IEEE for possible publication. Copyright may be transferred without notice, after which this version may no longer be accessible.}%
\thanks{Corresponding author: Qisong Wu.}%
\thanks{This work involved human subjects in its research. Approval of all ethical and experimental procedures and protocols was granted by the Special Committee for Science and Technology Ethics of Southeast University under Application No. SEU-EC-2026-0015. Informed consent was obtained from all participants.}%
\thanks{Jiuqi Yan, Chengbo Zhu, and Chengdong Xu are with the Key Laboratory of Underwater Acoustic Signal Processing of the Ministry of Education, Southeast University,
Nanjing 210096, China
(e-mail: 230248134@seu.edu.cn; 220241254@seu.edu.cn; 230228233@seu.edu.cn).}%
\thanks{Qisong Wu is with the Key Laboratory of Underwater Acoustic Signal Processing of the Ministry of Education, Southeast University, Nanjing 210096, China, and also with the Purple Mountain Laboratories, Nanjing 211111, China (e-mail: qisong.wu@seu.edu.cn).}
\endgraf}

\maketitle

\begin{abstract}
Millimeter-wave (mmWave) radar provides privacy-preserving human activity
recognition (HAR), but sparse point clouds limit its representation of body
structure and motion. We propose a cross-modal knowledge distillation
framework in which a teacher trained on camera-derived two-dimensional
(2D) human keypoints supervises a radar-only student. The student combines
PointNet-based point-set encoding, bidirectional temporal modeling,
segment-level aggregation, and
lightweight shared latent-channel scaling. The teacher retains contextualized
states at every time step, whereas the student summarizes the radar sequence
into a small number of segments, making fixed timestamp correspondence
unsuitable for feature transfer. Soft Temporal Alignment Distillation lets
each student segment adaptively retrieve and aggregate teacher temporal
evidence into its distillation target. Alignment denotes student-conditioned
representation-space matching rather than chronological or monotonic
registration. On the Multi-Modal Non-Intrusive 4D Human Dataset (MM-Fi),
the method improves radar-only accuracy from
87.04\% to 89.17\%, achieving the highest mean accuracy among the evaluated
established and recent distillation baselines. It also obtains higher mean
accuracy and macro-averaged F1 score (macro-F1) than the radar student under most evaluated point- and
frame-dropout settings. Adaptation to our independently collected
millimeter-wave Office Human Activity Recognition Dataset (mmO-HAR)
improves accuracy from 80.33\% to 87.75\%, supporting the
transferability of the distilled radar representation. The teacher and
alignment branch are removed after training, enabling privacy-preserving
radar-only inference without additional sensing or computational branches.
\end{abstract}


\begin{figure*}[t]
\centering
\includegraphics[width=0.85\textwidth]{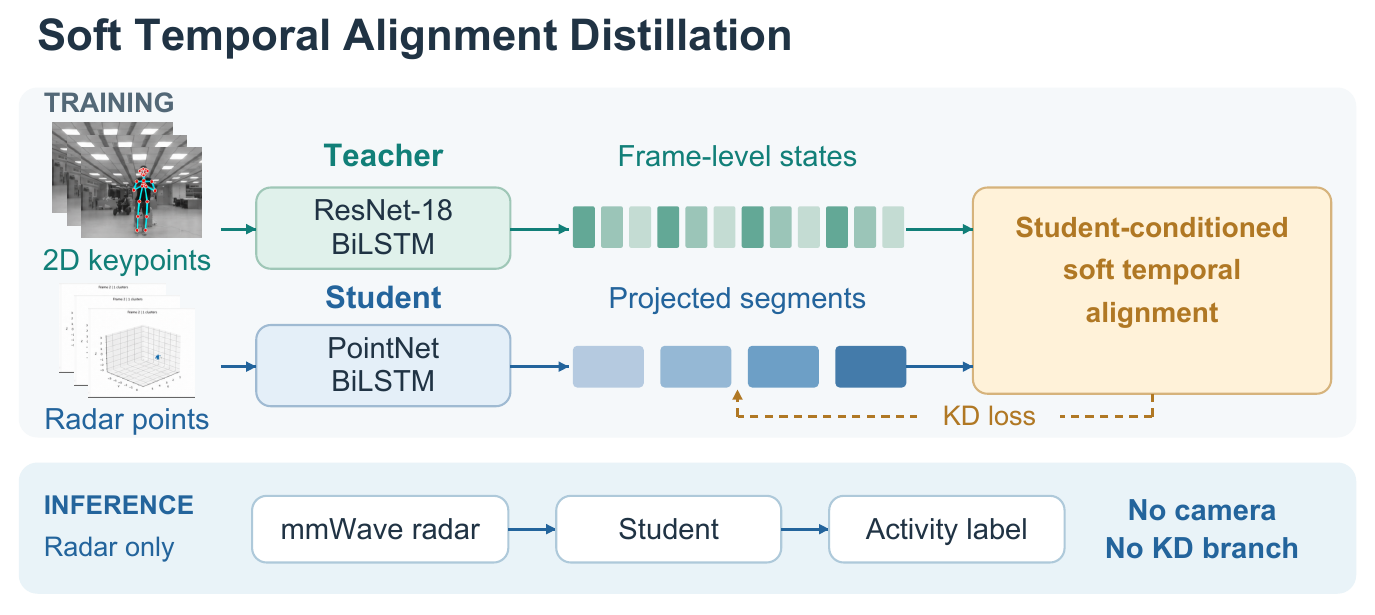}
\end{figure*}

\section{Introduction}

\IEEEPARstart{H}{uman} activity recognition (HAR) supports health monitoring
and assisted living \cite{florence2018medical, nicholson2024interventions}.
Contact-based systems use pressure sensors
\cite{agrawal2023fall, youngkong2021novel, brome2022fall}, gyroscopes
\cite{mukherjee2023novel, paramasivam2022internet, matos2023sensitivity},
and accelerometers
\cite{boudouane2023portable, lahouar2023development, abd2018fall}, but
physical contact and user burden motivate non-contact alternatives.

Cameras capture detailed appearance and motion. Camera-only HAR has used
diverse appearance and motion models
\cite{ros2023flexible,pranavan2023fall,paula2023novel,atikuzzaman2020human,xu2021action,lu2018deep}
as well as pose-based representations
\cite{liu2025systematic,liu2022representation}, but raises privacy
concerns and remains sensitive to illumination, background, and
occlusion. Camera-derived
keypoints \cite{zohirov2026human} retain body configuration and joint
trajectories in fixed-order two-dimensional (2D) coordinates while discarding most color,
texture, and identity-related appearance. These pose sequences can
supervise radar learning during training. Device-free radio-frequency
sensing offers unobtrusive, privacy-aware indoor HAR without wearable
sensors \cite{wang2022radar}.

We consider synchronized camera-derived keypoints and radar measurements
during model development, with only the millimeter-wave (mmWave) radar and compact student
used for inference. Removing the keypoint teacher and alignment branch
after training transfers camera-derived motion knowledge without requiring
a camera feed or teacher computation during operation.

mmWave radar operates under low-light conditions and
offers resilience to visual occlusion
\cite{newaz2023methods,qian2022parallel,wang2022radar,li2013review}.
Radar-only work has modeled spectral signatures and sparse point clouds
for activity recognition and pose estimation
\cite{he2023fall,li2023radar,shi2023robust,zhang2025lpbs,zhu2024mradhprs}.
Sparse, noisy reflections nevertheless make action-relevant body
configuration and motion difficult to recover.

Knowledge distillation (KD) \cite{hinton2015distilling} offers a way to
use the more explicit pose dynamics available from camera-derived keypoints to supervise a radar student. Applying KD in this setting,
however, involves three closely related challenges.

First, the teacher observes a fixed-size, explicitly indexed array of 2D
joint coordinates, whereas the student receives sparse, unordered three-dimensional (3D)
reflections augmented with Doppler and intensity measurements. Their
features therefore have no direct point-wise or feature-dimensional
correspondence.
Second, although the two streams are synchronized during acquisition, the
states that are most useful for recognition need not be emphasized at
identical temporal positions because keypoints and radar respond to motion
through different observation processes. Third, the teacher retains
contextualized states for all $T$ time steps, whereas the student
compresses the radar sequence into $K$ segment representations
($K\ll T$). A fixed frame-to-segment assignment would therefore impose a
rigid correspondence across both heterogeneous modalities and unequal
temporal resolutions.

Our framework addresses these challenges through a radar student with
PointNet~\cite{qi2017pointnet}, a permutation-invariant point-set encoder,
followed by a bidirectional long short-term memory (LSTM)
network~\cite{hochreiter1997lstm}, temporal segment pooling, and shared
latent-channel scaling. Soft Temporal Alignment Distillation (Soft-Align KD)
uses each projected student segment to query the complete teacher sequence
and construct an attention-weighted feature target. On the Multi-Modal
Non-Intrusive 4D Human Dataset (MM-Fi)~\cite{yang2023mm}, accuracy
increases from 87.04\% to 89.17\%, with higher mean scores under evaluated
point and frame dropout. Fine-tuning on the independently collected
millimeter-wave Office Human Activity Recognition Dataset (mmO-HAR)
further supports representation transferability.

The main contributions are summarized as follows:

1. We develop a radar-oriented student network that integrates
PointNet-based frame encoding, bidirectional temporal modeling, segment
pooling, and lightweight shared latent-channel scaling for sparse mmWave point-cloud sequences.

2. We propose Soft Temporal Alignment Distillation, which constructs a
separate teacher target for each student segment by adaptively aggregating contextualized keypoint states over the complete temporal sequence, avoiding a predetermined frame-to-segment assignment. The resulting matching is non-monotonic and is intended as representation-space evidence retrieval rather than chronological frame registration.

3. Experiments on MM-Fi evaluate recognition performance and robustness to point and frame dropout, while adaptation to our self-collected mmO-HAR dataset evaluates cross-dataset transferability under a subject-disjoint target-domain protocol. Complexity analysis further confirms that Soft Temporal Alignment Distillation introduces no additional parameters or computation at radar-only inference relative to the standalone student.

The implementation is publicly available at \url{https://github.com/BlakeYan97/Soft-Align-KD}.

Section II reviews related HAR and KD methods; Section III presents the
networks and distillation strategy; Section IV describes implementation and
evaluation protocols; Section V reports comparisons, ablations, robustness,
and adaptation experiments; and Section VI concludes the paper.

\section{Related Work}

\subsection{Multimodal Fusion with Radar}

Radar--visual fusion has been explored at several representation levels.
Kulhandjian et al.~\cite{kulhandjian2023ai} fused micro-Doppler radar and
infrared-camera observations for human detection in heavy smoke, reporting
98\% validation accuracy. For HAR, Jeon and Woo~\cite{jeon2026lightweight}
concatenated temporal radar range--Doppler and ultra-low-resolution camera
features, achieving 98.74\% accuracy on their 15-class dataset. The
multimodal fusion model M2FM~\cite{yi2025m2fm} instead combines decisions from a radar branch encoding
micro-Doppler and cadence--velocity features and a video Transformer branch
through Dempster--Shafer fusion; it reports 99.3\% accuracy on eight
self-collected actions. Yang et al.~\cite{yang2025crosstarget} concatenate
radar-spectrogram and video features and add intra- and cross-modal
semantic interaction with meta-prototype learning to improve recognition
of unseen subjects. X-Fi~\cite{chen2025xfi} uses cross-attention to retain
modality-specific information within a shared representation and supports
variable sensor combinations, including single-modality input, on MM-Fi
and the XRF55 multimodal sensing dataset. These results use different tasks and protocols and are not
directly ranked against ours. Our focus is training-only visual guidance
for a radar-only student, rather than multisensor feature or decision fusion.

\subsection{Cross-Modal Knowledge Distillation}

Cross-modal KD can retain multimodal guidance while simplifying the
student's sensing input. Fan et al.~\cite{fan2025cross} distill both
intermediate features and outputs from an audio-visual speech-separation
teacher into an audio-only student; their multi-stage teacher fusion
and distillation yield 15--25\% relative improvements in separation
metrics over the audio-only baseline. Wang et al.~\cite{wang2025uncertainty}
transfer color-and-depth (RGB-D) pose-estimation features to a depth-only student, weighting
distillation by learned uncertainty to emphasize reliable samples.
Park et al.~\cite{park2025cross} combine latent-feature and output-feature
transfer from a teacher combining light detection and ranging (LiDAR),
radar, Global Positioning System (GPS), and camera inputs to a radar-only beam
predictor, raising top-1 accuracy from 84.12\% to 86.33\% in their
simulated setting. Zhou et al.'s CIKD framework~\cite{zhou2025improving}
uses Contrastive Language--Image Pre-training (CLIP) features for
outlier weighting and text-feature-guided logits (unnormalized class scores) for image-classification
KD, where both task networks still process red--green--blue (RGB) images. These studies
address different tasks and transfer levels, not radar-segment to
keypoint-frame correspondence.

For heterogeneous sensors, RadarDistill~\cite{bang2024radardistill}
transfers LiDAR features to radar for 3D detection through cross-modality
alignment and activation- and proposal-based feature losses; it reports
20.5\% mean average precision (mAP) and a 43.7\% nuScenes detection score
(NDS) on nuScenes. CRKD~\cite{zhao2024crkd}
distills a LiDAR--camera detector into a camera--radar detector using
feature and response supervision in bird's-eye-view space, improving
nuScenes detection while retaining camera input at inference.
C$^2$KD~\cite{huo2024c2kd} addresses modality imbalance and mismatched
soft labels through on-the-fly selection, bidirectional distillation,
and proxy networks, improving transfer on audio-visual, image-text,
and RGB-depth tasks. These works target spatial correspondence,
prediction transfer, or modality gaps rather than unequal temporal
granularity in pose-to-radar HAR.

Recent general KD objectives provide complementary baselines. Logit
standardization for knowledge distillation (LS-KD)~\cite{sun2024logit}
standardizes teacher and student logits before temperature-scaled
Kullback--Leibler (KL) divergence matching, improving image-classification KD without
requiring their logit magnitudes to match. Logit-level Wasserstein knowledge
distillation (WKD-L)~\cite{lv2024wasserstein} uses discrete Wasserstein transport to transfer
inter-class relations and improves logit distillation on classification
and detection benchmarks. Both operate on global predictions rather
than constructing temporally localized teacher evidence.

Most closely, Shin et al.~\cite{shin2025graph} train a skeleton-based
spatial--temporal graph convolutional network (ST-GCN) teacher and radar
ST-GCN student with softened-response KL and
cosine feature-hint losses. They report accuracy gains of up to 7.46\%
on the Dongguk University Human Activity (DGUHA) dataset and 7.89\% on
MilliPoint and deploy a compressed student on a microcontroller unit (MCU).
Their hints directly match corresponding graph-network features;
our student instead compresses radar frames into segments and retrieves
a separate target for each segment from the complete teacher state
sequence. Tac-Mamba~\cite{electronics15071535} also uses pose guidance
for radar HAR: a pose teacher guides a radar student through topological
supervision and auxiliary pose regression, while trust-gated radar--pose
attention supports multimodal inference and a radar-only fallback. It
reports 87.54\% radar-only accuracy under an MM-Fi cross-environment
protocol and includes deployment on an NVIDIA Jetson Nano.
Unlike its pose-guided fusion and supervision, our approach performs
cross-modal knowledge distillation by retrieving frame-level teacher
evidence conditioned on each student segment. The key distinction is
the explicit modeling of temporal correspondence between frame-level
teacher states and segment-level student representations. The
original-paper results discussed above were obtained using different
datasets or evaluation protocols; common-protocol adaptations are
evaluated separately in
Section~\ref{sec:results_analysis}.

\section{Methodology}

\subsection{Problem Formulation}

Given a radar point cloud sequence
$\mathcal{P}=\{P_1,\ldots,P_T\}$ and a synchronized sequence of
camera-derived 2D human keypoints
$\mathcal{Q}=\{Q_1,\ldots,Q_T\}$, both containing $T$ frames, each paired sample is associated with an
action label $y\in\{1,\ldots,C\}$, where $C$ is the number of action
classes. The goal is to learn a radar-based student that predicts $y$ from
$\mathcal{P}$. During training, $\mathcal{Q}$ is available to a teacher and
provides cross-modal supervision to the student. During inference, the
teacher branch is removed and only $\mathcal{P}$ is required.

Superscripts $S$ and $T$ identify student and teacher quantities,
respectively; $\mathbb{R}$ denotes the real numbers. For a vector $v$,
$[v]_j$ denotes its $j$-th coordinate, and $(\cdot)^{\mathsf{T}}$
denotes transpose. A semicolon within brackets denotes feature
concatenation.

\subsection{Radar Point Cloud Representation}

MM-Fi provides synchronized camera observations and mmWave point clouds,
whose detections describe target locations and motion.

The $i$-th detection in radar frame $t$ has five attributes:

\begin{equation}
p_i^t = (x_i^t, y_i^t, z_i^t, d_i^t, I_i^t),
\end{equation}

where $(x_i^t, y_i^t, z_i^t)$ denotes 3D coordinates, $d_i^t$ the radial
Doppler velocity relative to the radar, and $I_i^t$ the reflected intensity.

The frame point cloud is

\begin{equation}
P_t = \{ p_i^t \mid i = 1,2,\dots,N_t \},
\end{equation}

where $N_t$ denotes the number of detected points in frame $t$.

An activity sample contains $T$ frames, giving the student input

\begin{equation}
\mathcal{P} = \{ P_1, P_2, \dots, P_T \}.
\end{equation}

\subsection{Student Network for Radar Point Clouds}

The student encodes the five point attributes $(x,y,z,d,I)$ into latent
features before temporal modeling; subsequent channel scaling acts on
these latent features.

\subsubsection{Frame-wise Point Cloud Encoding}

For each frame $P_t$, a PointNet-based encoder maps the unordered radar
points to a fixed-dimensional representation. The three spatial
coordinates $(x,y,z)$ are first processed by a learned $3\times3$ input
transformation. The transformed coordinates are then concatenated with
the original Doppler and intensity attributes, which are not included in
the spatial transformation. The resulting five-dimensional point features
are processed by shared $1\times1$ convolutional layers, which apply the
same learned channel mapping independently to every point, with channel
dimensions $5\rightarrow D_1\rightarrow D_2\rightarrow D_p$. Here,
$D_1$ and $D_2$ are intermediate channel widths, and $D_p$ is the
frame-feature dimension. Batch normalization, which normalizes channel
activations using mini-batch statistics during training, is applied after
each convolution. The rectified linear unit (ReLU), defined by
$\operatorname{ReLU}(a)=\max(a,0)$ for a scalar activation $a$, is used
after the first two convolutions, while the final layer is left without
ReLU before point-wise aggregation.

Following the first convolution, a learned $D_1\times D_1$ feature
transformation is applied to the intermediate point features. Both the
input- and feature-transformation subnetworks follow the PointNet design:
shared convolutional layers are followed by symmetric max pooling and
fully connected layers. Their layer widths are specified in
Table~\ref{tab:architecture_settings}.
Their final layers predict the corresponding transformation matrices,
with identity matrices added to the predictions. Batch normalization and
ReLU are applied to all intermediate layers of the transformation
subnetworks.

Finally, element-wise max pooling, which retains the largest response in
each channel across valid points, produces the
frame-level representation:

\begin{equation}
f_t^S = \text{PointNet}(P_t), \quad f_t^S \in \mathbb{R}^{D_p},
\end{equation}

where $D_p$ is the frame-feature dimension. The symmetric point-wise pooling makes the frame
representation invariant to the ordering of points within the frame.
Applying this encoder independently to all frames yields the temporal
feature sequence

\begin{equation}
F^S = \{f_1^S, f_2^S, \ldots, f_T^S\}.
\end{equation}

\subsubsection{Temporal Modeling}

To model temporal dependencies, the frame-wise feature sequence is
processed by a bidirectional LSTM (BiLSTM) with $L_S$ recurrent layers and
$H_S$ hidden units per direction. An LSTM uses gated memory states to
model dependencies over time; bidirectional processing combines forward
and backward context. No dropout is used inside the student LSTM.
Dropout randomly suppresses activations during training as a regularizer.
At each time step, the
forward and backward hidden states are concatenated:

\begin{equation}
h_t^S =
\left[
\overrightarrow{h_t^S};
\overleftarrow{h_t^S}
\right]
= \operatorname{BiLSTM}(F^S)_t,
\end{equation}

where $h_t^S \in \mathbb{R}^{D_S}$ is the output at frame $t$ after
processing the complete sequence $F^S$, and $D_S=2H_S$ is the
concatenated output dimension.

The resulting temporal feature sequence is

\begin{equation}
H^S = \{h_1^S, h_2^S, \ldots, h_T^S\}.
\end{equation}

\subsubsection{Segment-Level Feature Aggregation}

The temporal sequence is partitioned into $K$ contiguous, non-overlapping
segments $\{\mathcal{T}_k\}_{k=1}^{K}$ that follow the original temporal
order, where $K$ is the number of segments and $1\leq K\leq T$.
The first $K-1$ segments have length $\lfloor T/K \rfloor$, where
$\lfloor\cdot\rfloor$ denotes rounding down to an integer, and the last
segment contains the remaining frames.
For each segment, element-wise max pooling is applied along the temporal
dimension:

\begin{equation}
\left[u_k^S\right]_j =
\max_{t \in \mathcal{T}_k}
\left[h_t^S\right]_j,
\quad j=1,\ldots,D_S,
\end{equation}

where $u_k^S \in \mathbb{R}^{D_S}$ is the latent summary of the $k$-th
local temporal window. The collection of segment-level features is

\begin{equation}
U^S = \{u_1^S, u_2^S, \ldots, u_K^S\}.
\end{equation}

This operation retains the maximum activation of each latent dimension
within each local window and exposes multiple local summaries for
segment-level distillation. It does not preserve the frame order or
non-maximal responses within a segment.

\subsubsection{Shared Latent-Channel Scaling}

After segment-level aggregation, a lightweight channel-wise scaling
operation adjusts the contribution of the pooled latent dimensions.
Specifically, a learnable vector $r \in \mathbb{R}^{D_S}$, initialized to
an all-ones vector, is shared across all segments:

\begin{equation}
\bar{u}_k^S = r \odot u_k^S,
\end{equation}

where $\bar{u}_k^S$ is the scaled segment representation and $\odot$
denotes element-wise multiplication. The entries of
$u_k^S$ are hidden representations formed by jointly encoding the input
coordinates, Doppler velocity, intensity, and temporal context. Therefore,
the learned coefficients in $r$ do not correspond one-to-one to the five
physical input attributes. Instead, they rescale latent-channel
contributions, allowing channels with disproportionate responses to be
attenuated and informative channels to be emphasized before classification
and distillation. This operation does not reconstruct the temporal
information discarded by max pooling. Moreover, because $r$ is shared by
all samples and segments, it is a static learnable scaling rather than a
sample-dependent attention or gating mechanism.

\subsubsection{Action Classification and Distillation Projection}

For action classification, element-wise global max pooling is applied
across the $K$ scaled segment features to obtain the sequence-level
representation $g^S\in\mathbb{R}^{D_S}$:

\begin{equation}
\left[g^S\right]_j =
\max_{k=1,\ldots,K}
\left[\bar{u}_k^S\right]_j,
\quad j=1,\ldots,D_S.
\end{equation}

The classification logits are produced by a two-layer classifier
$\mathcal{C}_S$ with dimensions $D_S\rightarrow D_C\rightarrow C$:

\begin{equation}
o^S=\mathcal{C}_S(g^S),
\qquad
\hat{y}^S = \operatorname{Softmax}(o^S),
\end{equation}

where $o^S\in\mathbb{R}^{C}$ contains the unnormalized class scores
(logits), and $\hat{y}^S\in\mathbb{R}^{C}$ contains class probabilities.
Softmax exponentiates each logit and divides by the sum of exponentiated
logits, producing nonnegative probabilities that sum to one. The predicted
action is the class with the largest probability.
The $D_C$-dimensional intermediate layer is followed by batch
normalization, ReLU activation, and dropout before the final output layer.
Here, $C$ denotes the number of action classes.
Because max pooling across segments is invariant to their permutation,
the classification branch aggregates salient evidence from the local
windows but does not explicitly encode their order. The separate segment
representations remain available as local units for the segment-wise
distillation objective described below.

To match the dimensionality of the teacher temporal features, each scaled
student segment is additionally mapped to a teacher-compatible distillation
space:

\begin{equation}
z_k^S = W_A \bar{u}_k^S + b_A,
\quad z_k^S \in \mathbb{R}^{D_A},
\end{equation}

where $W_A \in \mathbb{R}^{D_A \times D_S}$ and
$b_A \in \mathbb{R}^{D_A}$ are learnable projection parameters. The
alignment dimension $D_A$ matches the output dimension of the teacher
bidirectional LSTM. The alignment embeddings
$Z^S=\{z_k^S\}_{k=1}^{K}$ are used only by the distillation mechanism, while
$g^S$ is used by the classification branch.

\begin{figure}[htb]
    \centering
    \includegraphics[scale=0.343]{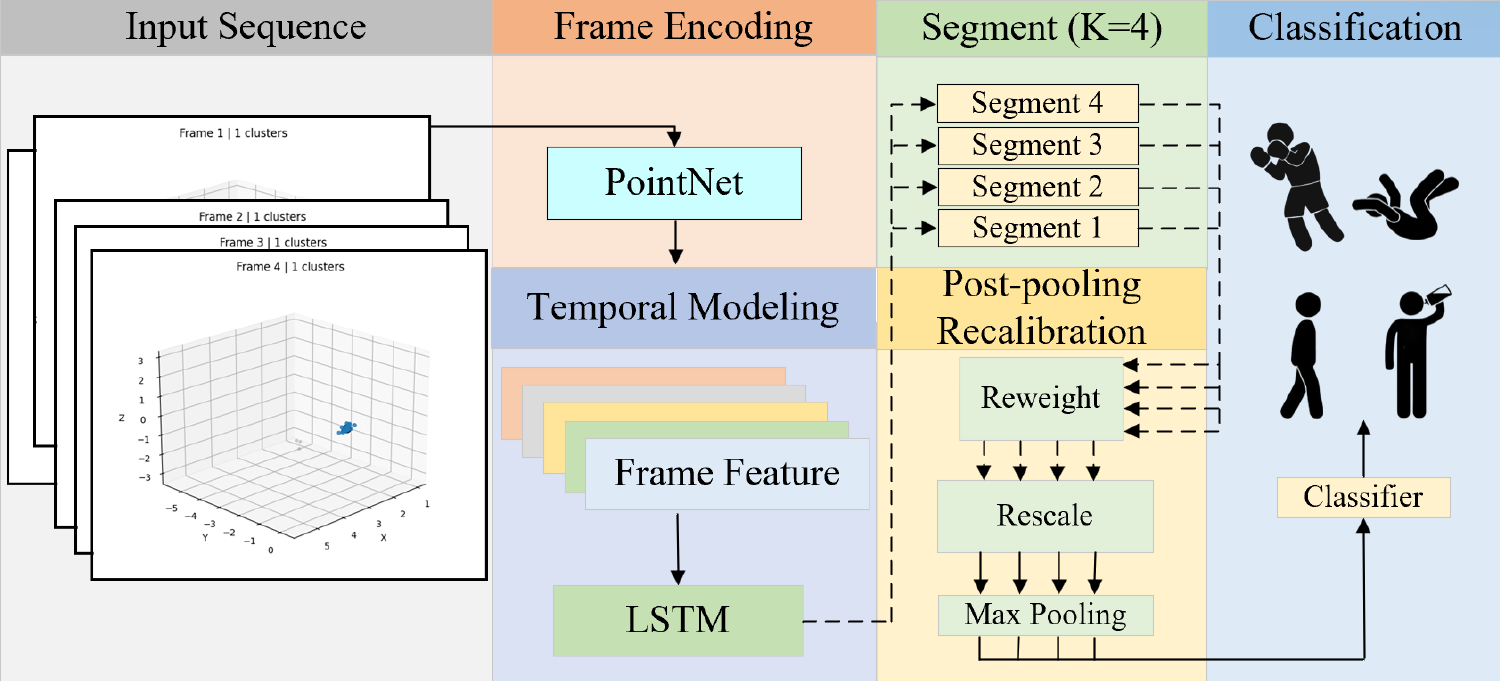}
    \caption{Architecture of the radar-based student network. It consists of a PointNet-based frame-wise encoder, a bidirectional LSTM for temporal modeling, temporal segment pooling, shared latent-channel scaling, and a classification head. Segment-level embeddings are additionally projected for knowledge distillation.}
    \label{fig:student_architecture}
\end{figure}

Fig.~\ref{fig:student_architecture} illustrates the structure of the student
network.

\subsection{Teacher Network for Camera-Derived 2D Keypoints}

The teacher operates on 2D human keypoints derived from the RGB camera
modality of MM-Fi \cite{yang2023mm}, rather than on raw RGB or grayscale
images. For each time step $t$, the camera-derived observation is
represented as

\begin{equation}
Q_t =
\left[
q_{t,1},q_{t,2},\ldots,q_{t,J}
\right]^{\mathsf{T}}
\in \mathbb{R}^{J\times2},
\quad
q_{t,j}=(u_{t,j},v_{t,j}),
\end{equation}

where $J$ is the number of body joints, and $(u_{t,j},v_{t,j})$ denotes
the image-plane coordinates of the $j$-th joint in frame $t$.
The $J$ joints follow a fixed order, and the $T$
frames are synchronized with the radar sequence. The teacher input
is therefore

\begin{equation}
\mathcal{Q}=\{Q_1,Q_2,\ldots,Q_T\}.
\end{equation}

Keypoints retain body configuration and joint trajectories while
suppressing color, texture, background, and identity-related appearance,
focusing teacher supervision on pose transitions and motion dynamics.

\subsubsection{Frame-wise Keypoint Encoding}

To process each keypoint frame with a convolutional backbone, $Q_t$ is
treated as a single-channel coordinate tensor of size $1\times J\times2$.
We employ an 18-layer residual network (ResNet-18)~\cite{he2016resnet},
whose skip connections add a shortcut of each block's input to its learned residual mapping
to facilitate optimization. Its first convolution is replaced by a
single-channel layer. The remaining backbone follows the standard
ResNet-18 configuration. Global average pooling, which averages each
feature channel over its spatial positions, produces a
$D_F$-dimensional frame-level feature:

\begin{equation}
f_t^T = \operatorname{ResNet18}(Q_t),
\quad
f_t^T \in \mathbb{R}^{D_F}.
\end{equation}

The ResNet-18 backbone is initialized with ImageNet-pretrained parameters,
except for the newly constructed single-channel input convolution, which
is randomly initialized. This convolutional neural network (CNN) backbone is subsequently fine-tuned
end-to-end on MM-Fi together with the temporal encoder and classifier.
Applying the frame encoder independently across the sequence gives

\begin{equation}
F^T = \{f_1^T,f_2^T,\ldots,f_T^T\}.
\end{equation}

\subsubsection{Temporal Modeling}

The frame-level feature sequence is processed by a bidirectional LSTM
with $L_T$ recurrent layers and $H_T$ hidden units per direction. Dropout is
applied between the two recurrent layers. The
forward and backward states at each time step are concatenated:

\begin{equation}
h_t^T =
\left[
\overrightarrow{h_t^T};
\overleftarrow{h_t^T}
\right]
= \operatorname{BiLSTM}(F^T)_t,
\quad
h_t^T \in \mathbb{R}^{D_A},
\end{equation}

where the subscript $t$ selects the output at frame $t$ after processing
the complete sequence $F^T$, and $D_A=2H_T$. This dimension is also used as the common
alignment dimension in the distillation stage. The complete teacher
temporal sequence is denoted by

\begin{equation}
H^T=\{h_1^T,h_2^T,\ldots,h_T^T\}.
\end{equation}

\subsubsection{Temporal Feature Aggregation and Classification}

For teacher classification, element-wise max pooling is applied over the
temporal dimension:

\begin{equation}
\left[g^T\right]_j =
\max_{t=1,\ldots,T}
\left[h_t^T\right]_j,
\quad j=1,\ldots,D_A.
\end{equation}

The resulting sequence-level representation is mapped to classification
logits and class probabilities by a linear classifier:

\begin{equation}
o^T=W_Tg^T+b_T,
\qquad
\hat{y}^T = \operatorname{Softmax}(o^T),
\end{equation}

where $W_T\in\mathbb{R}^{C\times D_A}$ and
$b_T\in\mathbb{R}^{C}$ are the learned classifier weights and bias;
$o^T$ and $\hat{y}^T$ are the teacher logits and class probabilities.
The teacher is first trained independently using
the action classification objective. During student training, its
parameters are frozen, and the unpooled temporal sequence $H^T$, rather
than the pooled classification feature $g^T$, is retained to provide
contextualized temporal states for distillation.

\begin{figure*}[htp]
    \centering
    \includegraphics[width=\textwidth]{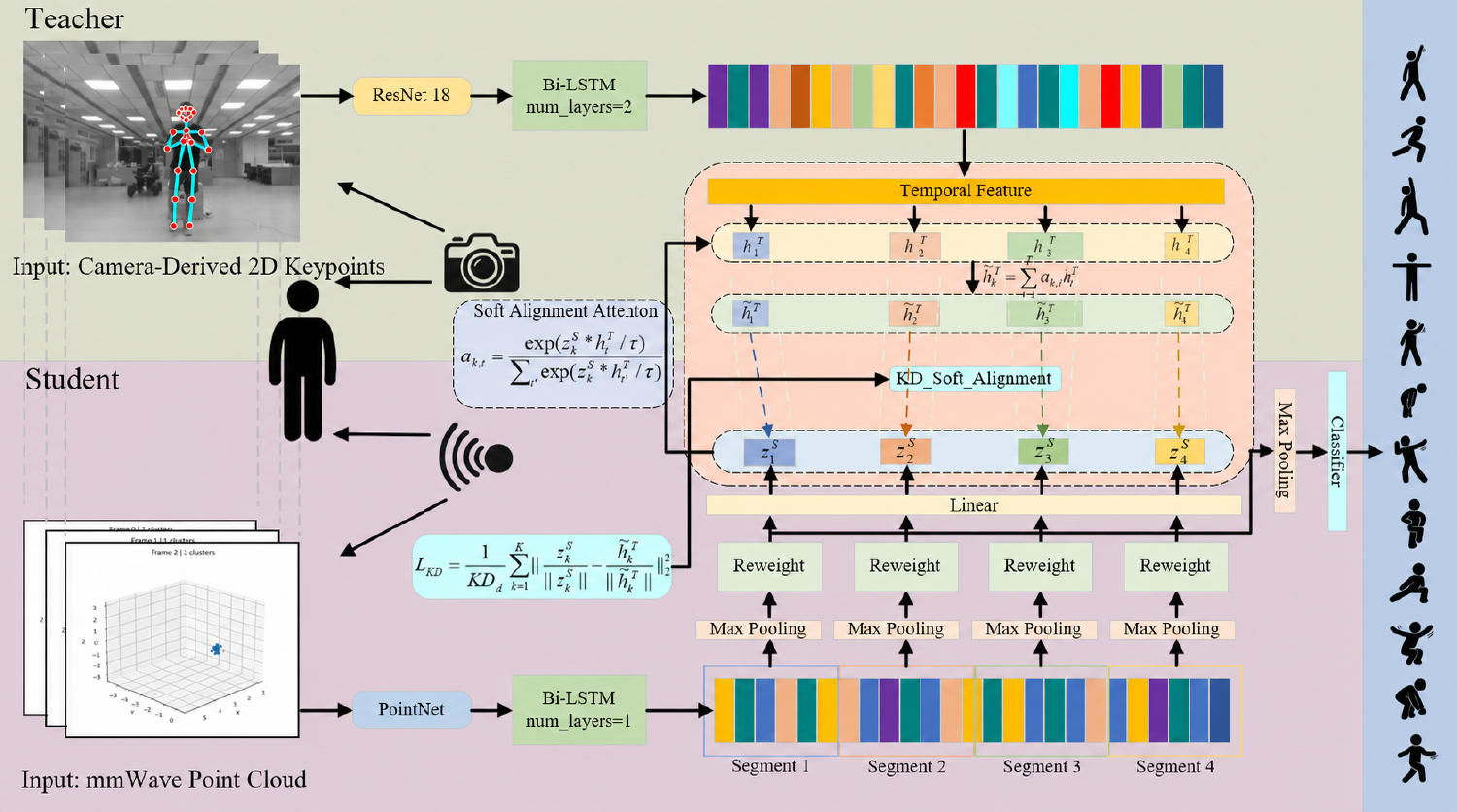}
    \caption{Overview of the proposed soft temporal alignment distillation. The teacher network (ResNet-18+BiLSTM) processes camera-derived 2D joint-coordinate tensors to produce frame-level temporal features $\{h_t^T\}_{t=1}^{T}$. The photograph illustrates keypoint acquisition; the teacher receives joint coordinates, not the image. The student network processes radar point clouds to produce projected segment-level alignment embeddings $\{z_k^S\}_{k=1}^{K}$. For each student segment, attention weights $\alpha_{k,t}$ are computed over the teacher temporal sequence, and the weighted teacher features form the distillation target.}
    \label{fig:soft_kd}
\end{figure*} 

\subsection{Soft Temporal Alignment Distillation}

Distillation must bridge indexed 2D joint trajectories and sparse,
unordered radar reflections, as well as $T$ teacher states and $K$ student
segments ($K\ll T$). Acquisition synchronization does not ensure that
discriminative responses coincide; predetermined frame-to-segment matching
would impose a rigid correspondence across unequal temporal granularities.

We instead retrieve a separate teacher target for each student segment.
The frozen teacher supplies $H^T=\{h_t^T\}_{t=1}^{T}$, and the student
projection supplies queries $Z^S=\{z_k^S\}_{k=1}^{K}$ of the same dimension
$D_A$. Dimensional compatibility alone does not establish semantic
correspondence: the student encoder and projection learn retrieval jointly
under classification and distillation objectives.

\subsubsection{Student-Conditioned Temporal Evidence Retrieval}

For the $k$-th student segment, attention weights over all $T$ teacher
states are obtained by applying a temporal softmax to temperature-scaled
dot-product similarities:

\begin{equation}
\alpha_{k,t} =
\frac{
\exp\left(\left(z_k^S\right)^{\mathsf{T}}h_t^T/\tau\right)}
{\sum_{t'=1}^{T}
\exp\left(\left(z_k^S\right)^{\mathsf{T}}h_{t'}^T/\tau\right)},
\qquad
\sum_{t=1}^{T}\alpha_{k,t}=1,
\end{equation}

where the attention temperature $\tau>0$ controls concentration, $z_k^S$ is the query,
and teacher states are both keys and values. The weight $\alpha_{k,t}$
measures teacher state $t$'s contribution to radar segment $k$'s target.
Thus, alignment constructs a relevant feature-space target rather than
estimating a physical frame correspondence or temporal warping path.

The segment-conditioned teacher target is the weighted sum

\begin{equation}
\tilde{h}_k^T =
\sum_{t=1}^{T}
\alpha_{k,t} h_t^T.
\end{equation}

Since $\alpha_{k,t}\geq0$ and the weights sum to one, $\tilde{h}_k^T$
convexly aggregates evidence across teacher time steps.

Independent normalization imposes neither exclusivity nor monotonicity:
segments may share discriminative teacher evidence, and attention peaks
need not follow temporal order. This permits many-to-one retrieval when
useful for classification. The ablation analyses quantify these patterns
and examine an order-and-diversity constrained variant.

\subsubsection{Feature Distillation Loss}

To transfer the retrieved teacher evidence without forcing the two
modalities to match feature magnitudes, both representations are
$\ell_2$-normalized before computing the element-wise mean-squared error (MSE):

\begin{equation}
\mathcal{L}_{KD} =
\frac{1}{K D_A}
\sum_{k=1}^{K}
\left\|
\operatorname{norm}_{\epsilon}(z_k^S)
-
\operatorname{norm}_{\epsilon}(\tilde{h}_k^T)
\right\|_2^2 .
\end{equation}

The factor $D_A$ gives the element-wise mean-squared error, which is then
averaged over the mini-batch. Here,
$\operatorname{norm}_{\epsilon}(v)=v/\max(\|v\|_2,\epsilon)$, where
$v$ is a feature vector, $\|v\|_2$ is its Euclidean norm, and
$\epsilon>0$ is a fixed numerical safeguard, not a tuned hyperparameter.
Fig.~\ref{fig:soft_kd} omits this safeguard; its ordinary $\ell_2$
normalization is identical when $\|v\|_2>\epsilon$. Normalization primarily
compares representation directions rather than magnitudes.

The teacher and $H^T$ remain fixed, but attention is differentiable with
respect to $z_k^S$. Gradients from $\mathcal{L}_{KD}$ update the student
encoder and projection, jointly learning evidence retrieval and normalized
feature matching.


\subsubsection{Overall Training Objective}

The student combines ground-truth classification and feature distillation:

\begin{equation}
\mathcal{L}
=
\mathcal{L}_{CE}
+
\lambda
\mathcal{L}_{KD},
\end{equation}

where $\mathcal{L}_{CE}$ is the cross-entropy classification loss, equal to
the negative logarithm of the probability assigned to the true action
label $y$, and $\lambda\geq0$ weights distillation. The teacher supplies
unpooled temporal states, not logits, and is removed for radar-only
inference. Fig.~\ref{fig:soft_kd} summarizes the framework.

\section{Implementation and Experimental Details}

\subsection{Experimental Databases}

The main experiments use the public \textit{MM-Fi} dataset
\cite{yang2023mm}, comprising 40 subjects, 27 activities, and four indoor
environments (E01--E04), with synchronized RGB, depth, LiDAR, Wi-Fi
channel state information (CSI),
and mmWave radar measurements. We use 17 camera-derived image-plane joints
per frame, rather than raw RGB, and radar detections containing 3D
coordinates, Doppler velocity, and intensity. Each action sample contains
297 synchronized frames. Keypoints supervise training only; radar is the
inference modality. Table~\ref{tab:mmfi_summary} summarizes the data used.

\begin{table}[h]
\centering
\caption{MM-Fi Dataset Summary.}
\label{tab:mmfi_summary}
\footnotesize
\setlength{\tabcolsep}{3pt}
\renewcommand{\arraystretch}{1.08}
\begin{tabularx}{\columnwidth}{>{\centering\arraybackslash}X>{\centering\arraybackslash}X}
\hline
\textbf{Property} & \textbf{Value} \\
\hline
Number of Subjects & 40 \\
Number of Actions & 27 \\
Number of Scenes & 4 \\
Modalities Used & Camera-derived 2D keypoints, mmWave radar \\
Evaluation Protocol & Cross-Subject \\
\hline
\end{tabularx}
\end{table}

\subsection{Data Splits and Evaluation Protocol}

Eight MM-Fi subjects (S05, S10, S15, S20, S25, S30, S35, and S40) form a
fixed test set. The other 32 form five subject-disjoint development folds;
each run uses one fold for validation and the others for optimization.
Test subjects are excluded from learning and checkpoint selection.

The highest-validation-accuracy checkpoint is evaluated on the fixed test
set. Accuracy is the fraction of correctly classified sequences.
The macro-averaged F1 score (macro-F1) is the unweighted mean of class-wise
F1 scores; each F1 score is the harmonic mean of precision (the fraction
of predictions for a class that are correct) and recall (the fraction of
true class members correctly identified).
Tables report the unweighted mean and standard deviation of accuracy
and macro-F1 across the five development-fold runs. All baselines and
ablations use the same folds and selection rule. The standard deviation
therefore describes run-to-run variability under this protocol, not
variation across five independent test folds.

\subsection{Implementation Details}

The framework uses PyTorch on an NVIDIA A100 graphics processing unit
(GPU). For MM-Fi, $T=297$ and $J=17$; $B$ denotes the mini-batch size.
The provided NumPy joint arrays are stacked into $(B,T,1,J,2)$ teacher tensors, without RGB
decoding, grayscale conversion, pose estimation, or keypoint rasterization.

Radar frames are adjusted to $N=112$ points by uniform subsampling without
replacement when at least 112 detections are available, or zero-padding
otherwise; empty frames contain only zeros. A point-validity mask
distinguishes valid radar detections from zero-padded entries and excludes
padded points from point-wise max pooling. The $(B,T,N,5)$ tensor is
permuted to $(B,T,5,N)$ for PointNet.
Both modalities use 32-bit floating point (FP32) tensors, with no additional
coordinate normalization or data augmentation.

Table~\ref{tab:architecture_settings} collects the numerical network
settings corresponding to the symbols in Section~III. The fixed
normalization safeguard in the distillation loss is
$\epsilon=10^{-12}$, as in PyTorch's \texttt{F.normalize}.

\begin{table}[t]
\centering
\caption{Network Architecture Settings.}
\label{tab:architecture_settings}
\footnotesize
\setlength{\tabcolsep}{3pt}
\renewcommand{\arraystretch}{1.08}
\begin{tabularx}{\columnwidth}{>{\raggedright\arraybackslash}p{0.27\columnwidth}>{\raggedright\arraybackslash}X}
\hline
\textbf{Component} & \textbf{Configuration} \\
\hline
Student point encoder & Shared convolutions $5\rightarrow D_1\rightarrow D_2\rightarrow D_p$, with $D_1=64$, $D_2=128$, and $D_p=1024$. \\
PointNet transforms & Input $3\times3$; feature $D_1\times D_1=64\times64$. Each transformation subnetwork has convolutional output widths $64\rightarrow128\rightarrow1024$ and fully connected widths $1024\rightarrow512\rightarrow256$. \\
Student temporal encoder & $L_S=1$ bidirectional LSTM layer; $H_S=256$ units per direction and $D_S=2H_S=512$; no LSTM dropout. \\
Student classifier & $D_S\rightarrow D_C\rightarrow C$, with $D_C=256$ and $C=27$; dropout rate $0.3$. \\
Teacher frame encoder & ResNet-18: first convolution $1\rightarrow64$, $7\times7$ kernel, stride $2$, padding $3$; four two-block residual stages with output widths $64,128,256,512$; $D_F=512$. \\
Teacher temporal encoder & $L_T=2$ bidirectional LSTM layers; $H_T=512$ units per direction and $D_A=2H_T=1024$; inter-layer dropout rate $0.3$. \\
\hline
\end{tabularx}
\end{table}

Each fold's teacher is trained with cross-entropy for 50 epochs using
the adaptive moment estimation (Adam) optimizer, learning rate $1\times10^{-4}$, and weight decay
$1\times10^{-5}$. ImageNet-initialized ResNet-18 layers, the new
single-channel input convolution, two-layer BiLSTM, and classifier are
optimized jointly. The highest-validation-accuracy teacher is frozen for
the corresponding student run.

The student uses Adam, learning rate $1\times10^{-4}$, and 50 epochs,
with batch sizes of 32 for training and 16 for validation and testing.
There is no learning-rate scheduler or early stopping; validation accuracy
selects the best checkpoint. Cross-entropy and normalized feature MSE
follow Section~III, with $\lambda=0.5$, $\tau=0.08$, and $K=4$ unless
specified otherwise. Baselines and ablations preserve the optimizer,
splits, and selection procedure, changing only the investigated component
or objective.

For the recent logit-based baselines, we use their method-specific
recommended settings within the same training protocol. LS-KD uses a
temperature of 2, a distillation weight of 9, and a numerical
standardization constant of $10^{-7}$. WKD-L uses a temperature of 2, an
initial Wasserstein weight of 30 with cosine decay over the 50 training
epochs, and 10 Sinkhorn matrix-scaling iterations with entropic regularization 0.05. For
each fold, its class-cost matrix is computed only from the frozen teacher
logits of that fold's training subjects using linear centered kernel
alignment (CKA) and a sharpening coefficient of 1; validation and test
samples are not used to construct the matrix.

For deployment profiling, we use a batch size of one and a complete
297-frame MM-Fi sequence with 112 radar points per frame. The frame-wise
PointNet computations are vectorized over the folded batch--time axis.
Parameter counts and multiply--accumulate operations (MACs) are obtained
with PyTorch and the THOP model-profiling package, respectively. One
GMAC denotes $10^9$ MACs. GPU latency is averaged over 100 runs
after 10 warm-up runs on an NVIDIA A100-SXM4-40GB using PyTorch 2.6.0 and
the NVIDIA CUDA 12.4 GPU computing toolkit. The training-only alignment projection is removed from the
deployment student because it does not contribute to classification logits.

\subsection{mmO-HAR Dataset and Cross-Dataset Adaptation Protocol}
\label{sec:mmO-HAR}

For cross-environment and cross-sensor adaptation, we collect
\textbf{mmO-HAR} (millimeter-wave Office Human Activity Recognition
Dataset) using a Texas Instruments (TI) IWR6843ISK radar configured with the
mmWave Demo Visualizer and software development kit (SDK) version 3.6.
The radar uses three transmit and four
receive antennas, a 60-GHz start frequency, a 3.6817-GHz sweep bandwidth,
and a 40-MHz/$\mu$s chirp slope. Each chirp contains 192 analog-to-digital
converter (ADC) samples at
2.086~MS/s, and each frame contains 96 chirps. At 15 frames/s, the configured
range and radial-velocity resolutions are 0.041~m and 0.13~m/s, respectively;
the maximum unambiguous range and radial velocity are 6.3~m and 2~m/s.
Platform-generated point-cloud detections are exported without further
offline signal processing.

\begin{figure}[htbp]
    \centering
    \subfloat[IWR6843ISK\label{fig:device}]{%
        \includegraphics[width=0.48\linewidth]{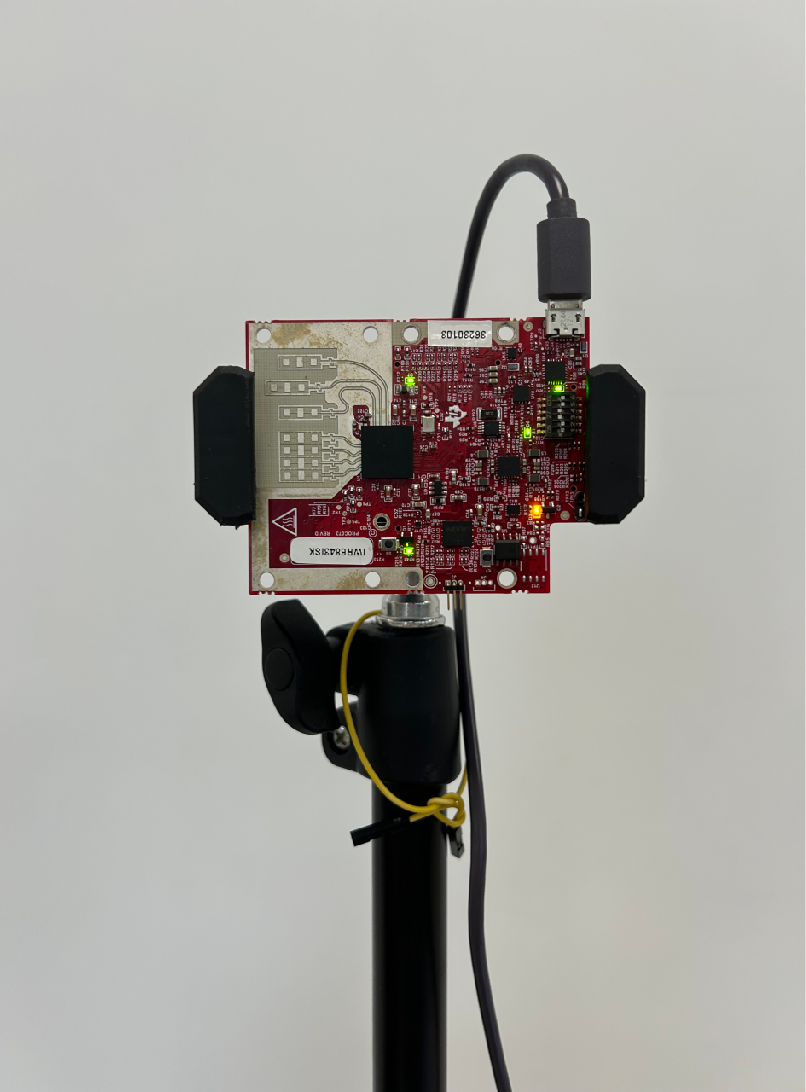}}
    \hfill
    \subfloat[Subject\label{fig:subject}]{%
        \includegraphics[width=0.48\linewidth]{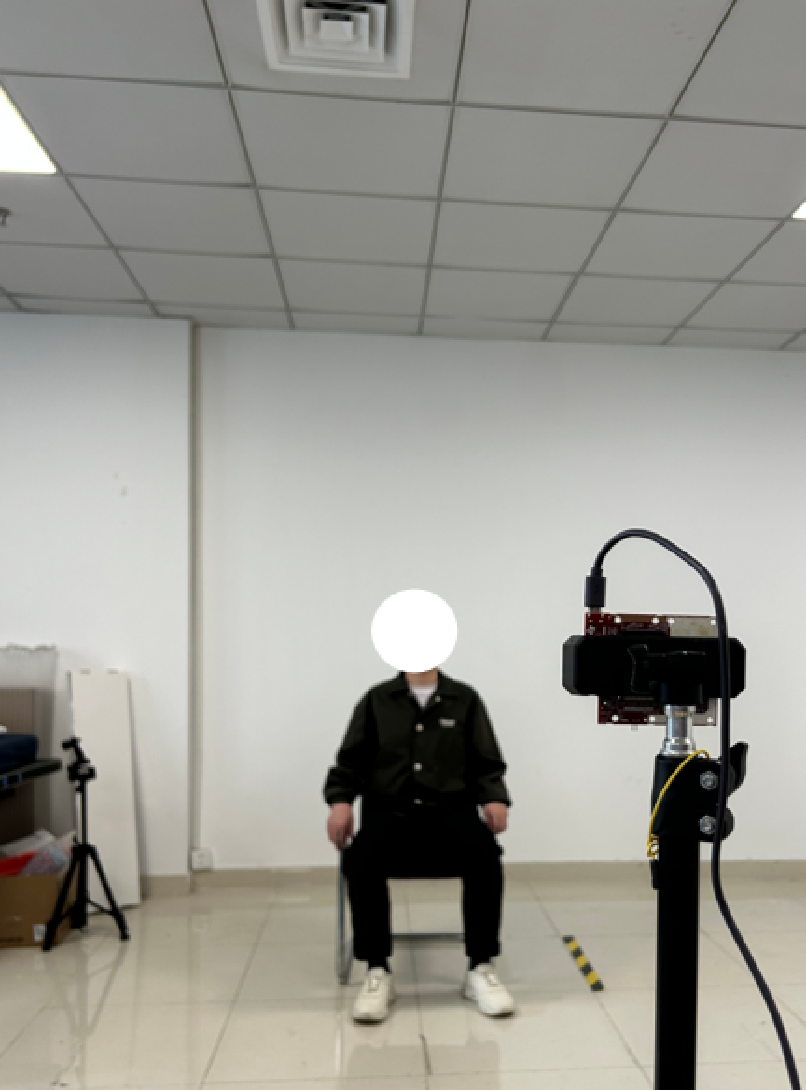}}
    \caption{Data collection setup for the mmO-HAR dataset. (a) Texas Instruments IWR6843ISK mmWave radar placed at a height of 1.2~m. (b) A subject performing a sitting down action at a distance of 3~m.}
    \label{fig:mmO-HAR}
\end{figure}

The radar is placed at a height of 1.2~m, and the subject stands 3~m
away, facing the radar. The data collection setup is illustrated in Fig.~\ref{fig:mmO-HAR}. The dataset contains recordings from eight subjects (S01--S08), each performing six types of actions: sitting down
(A01), standing up (A02), waving right hand (A03), waving left hand (A04), picking up an object (A05), and walking (A06). Each action lasts for 4 seconds (60 frames per sequence). Each subject repeats each action 20 times, resulting in a total of $8 \times 6 \times 20 = 960$ valid sequences.  

mmO-HAR is used only as the fine-tuning target in
Section~\ref{sec:generalization}, not in the main MM-Fi experiments.
Subjects S02, S03, S05, S06, S07, and S08 form the development set;
S01 and S04 form the subject-disjoint held-out test set. Five-fold
cross-validation on development sequences selects checkpoints for test
evaluation. Transfer initializes layers whose names and tensor shapes
match the MM-Fi student; the target output layer is newly initialized
for the different class count.

\section{Experimental Results and Analysis}

We evaluate MM-Fi performance, deployment complexity, architecture and
distillation ablations, corruption robustness, and mmO-HAR adaptation.
In the accuracy and macro-F1 tables, both means and standard deviations
are expressed on a 0--100 percentage scale.

\subsection{Results Analysis}
\label{sec:results_analysis}

Table~\ref{tab:main_results} compares the keypoint teacher, radar student,
and distilled student. The teacher achieves $(98.15\pm1.24)\%$ accuracy
and $(98.13\pm1.23)\%$ macro-F1, establishing the supervision source's
performance with explicit joint trajectories. Soft-Align KD raises the
student's accuracy from $(87.04\pm2.97)\%$ to $(89.17\pm6.71)\%$
and macro-F1 from $(87.10\pm2.75)\%$ to $(89.18\pm6.68)\%$: a
2.13-percentage-point accuracy gain without changing radar-only inference.
Its larger dispersion indicates sensitivity to the development protocol.
With identical folds and checkpoint selection, we report a mean gain,
not uniform fold-wise superiority or statistical significance, and retain
the standard deviations to expose this variability.

\begin{table}[h]
\centering
\caption{Recognition Performance on MM-Fi.}
\label{tab:main_results}
\footnotesize
\setlength{\tabcolsep}{3pt}
\renewcommand{\arraystretch}{1.08}
\begin{tabularx}{\columnwidth}{@{}>{\raggedright\arraybackslash}Xcc@{}}
\hline
Model & Accuracy (\%) & Macro-F1 (\%) \\
\hline
Teacher & $98.15 \pm 1.24$ & $98.13 \pm 1.23$ \\
Student & $87.04 \pm 2.97$ & $87.10 \pm 2.75$ \\ \hline
Student + Soft-Align KD & $\mathbf{89.17 \pm 6.71}$ & $\mathbf{89.18 \pm 6.68}$ \\
\hline
\end{tabularx}
\end{table}

We next compare Soft-Align KD with conventional objectives, LS-KD
\cite{sun2024logit}, WKD-L \cite{lv2024wasserstein}, and adaptations of
Shin et al.~\cite{shin2025graph} and Tac-Mamba~\cite{electronics15071535}
in Table~\ref{tab:kd_methods}. All use the same MM-Fi subject splits and
radar-only test protocol. Shin's response-and-hint adaptation retains our
student network; the Tac-Mamba adaptation retains our PointNet--BiLSTM core
but adds gated fusion and 2D-pose supervision. Neither reproduces the
original paper's full architecture.

\begin{table}[h]
\centering
\caption{Distillation Comparison on MM-Fi.}
\label{tab:kd_methods}
\footnotesize
\setlength{\tabcolsep}{3pt}
\renewcommand{\arraystretch}{1.08}
\begin{tabularx}{\columnwidth}{@{}>{\raggedright\arraybackslash}Xcc@{}}
\hline
KD Method & Accuracy (\%) & Macro-F1 (\%) \\
\hline
Logits KD \cite{hinton2015distilling} & $84.81 \pm 2.84$ & $84.61 \pm 3.03$ \\
LS-KD \cite{sun2024logit} & $84.54 \pm 5.84$ & $84.32 \pm 6.07$ \\
WKD-L \cite{lv2024wasserstein} & $88.15 \pm 5.66$ & $88.31 \pm 5.38$ \\ \hline
Feature KD \cite{romero2014fitnets} & $86.57 \pm 6.06$ & $86.47 \pm 5.99$ \\
Attention Transfer (AT) \cite{zagoruyko2016paying} & $86.57 \pm 4.90$ & $86.49 \pm 4.90$ \\
Feature + AT & $85.37 \pm 5.13$ & $85.26 \pm 4.93$ \\ \hline
Shin et al. (adapted) \cite{shin2025graph} & $87.96 \pm 4.51$ & $87.77 \pm 4.65$ \\
Tac-Mamba (adapted) \cite{electronics15071535} & $79.07 \pm 5.99$ & $78.95 \pm 5.84$ \\ \hline
Soft-Align KD (Ours) & $\mathbf{89.17 \pm 6.71}$ & $\mathbf{89.18 \pm 6.68}$ \\
\hline
\end{tabularx}
\end{table}

\begin{table*}[t]
\centering
\caption{Model Complexity and Latency.}
\label{tab:deployment_complexity}
\footnotesize
\setlength{\tabcolsep}{5pt}
\renewcommand{\arraystretch}{1.08}
\begin{tabular*}{\textwidth}{@{\extracolsep{\fill}}llrrrr@{}}
\hline
Model & Role & Params (M) & GMACs & FP32 size (MiB) & Latency (ms) \\
\hline
Teacher & Training only & 21.700 & 7.567 & 82.78 & $28.04\pm7.01$ \\
Radar student & Inference & 5.568 & 16.020 & 21.24 & $10.60\pm3.89$ \\
Soft-Align KD student & Inference & 5.568 & 16.020 & 21.24 & $10.60\pm3.89$ \\
\hline
\end{tabular*}
\end{table*}

Conventional objectives and LS-KD do not improve on the radar student
in this evaluation. Logit standardization alone provides no numerical gain.
Feature KD and AT are closer to, but below, the student;
their combination adds no gain. Shin's adapted response-and-hint objective
reaches 87.96\% accuracy, while the Tac-Mamba-inspired fusion and 2D-pose
adaptation reaches 79.07\%. The latter result characterizes this modified
implementation, not the original Mamba-based model. Soft-Align KD has the
highest mean, although its 1.21-point accuracy margin over Shin's adaptation
is small relative to the fold-wise dispersion.

WKD-L is the strongest competing baseline, with $(88.15\pm5.66)\%$
accuracy and $(88.31\pm5.38)\%$ macro-F1, gains of 1.11 and 1.21
percentage points over the student. This supports class-level transport
in the heterogeneous setting. Soft-Align KD's mean scores are higher by
1.02 and 0.87 points, respectively. Overlapping standard-deviation ranges
and the absence of paired hypothesis tests limit these to numerical,
not statistically significant, improvements. Unlike global logit targets,
segment-conditioned retrieval provides complementary temporal supervision
without assuming hard frame correspondence.

\subsection{Deployment Complexity}

Table~\ref{tab:deployment_complexity} reports model complexity for one
complete MM-Fi sequence. The deployed radar student contains 5.568~M
parameters and occupies 21.24~mebibytes (MiB) in FP32, reducing both quantities by
74.3\% relative to the training-only teacher. Its 16.020~GMAC count is
higher than the teacher's because the PointNet encoder is applied to every
radar frame; accordingly, operation count and measured latency should not
be interpreted as interchangeable measures. With vectorized frame encoding,
the student processes a sequence in $10.60\pm3.89$~ms on the specified GPU.
The independently trained student and the Soft-Align KD student have
identical deployment architectures and costs. The alignment projection adds
0.525~M parameters only during KD training, while the two explicit
Soft-Align matrix multiplications add approximately 0.00243~GMAC per sample.
Thus, distillation introduces no additional inference-time parameters,
operations, or sensing modalities. These measurements characterize server
hardware; embedded-device latency and energy consumption remain to be
evaluated separately.

\subsection{Ablation Studies}

We progressively add segment pooling and latent-channel scaling, vary
$K$ in the complete student, compare soft and fixed segment alignment,
and test an order-and-diversity constraint on attention behavior.

\begin{table}[h]
\centering
\caption{Student Architecture and Segment-Count Ablations.}
\label{tab:student_ablation}
\footnotesize
\setlength{\tabcolsep}{3pt}
\renewcommand{\arraystretch}{1.08}
\begin{tabularx}{\columnwidth}{@{}>{\raggedright\arraybackslash}Xcc@{}}
\hline
Model & Accuracy (\%) & Macro-F1 (\%) \\
\hline
PointNet + BiLSTM (Baseline) & $86.02 \pm 5.08$ & $86.06 \pm 4.97$ \\
+ Temporal Segment Pooling ($K=4$) & $86.30 \pm 3.85$ & $86.13 \pm 3.76$ \\
+ Segment + Reweight ($K=3$) & $82.78 \pm 3.43$ & $82.67 \pm 3.09$ \\
+ Segment + Reweight ($K=4$) & $\mathbf{87.04 \pm 2.97}$ & $\mathbf{87.10 \pm 2.75}$ \\
+ Segment + Reweight ($K=5$) & $83.70 \pm 2.92$ & $83.41 \pm 2.92$ \\
+ Segment + Reweight ($K=6$) & $81.02 \pm 4.31$ & $80.98 \pm 4.04$ \\
\hline
\end{tabularx}
\end{table}

\subsubsection{Role of Temporal Segment Pooling}

The baseline globally max-pools the BiLSTM sequence; segment pooling
instead exposes contiguous local summaries for distillation, although
classification pooling remains order-insensitive. Without an intervening
segment-dependent transformation, nested element-wise maxima satisfy
$\max_k\max_{t\in\mathcal{T}_k}[h_t^S]_j=\max_t[h_t^S]_j$; segment pooling
alone is algebraically equivalent to global max pooling for classification.
The small difference between the first two rows of
Table~\ref{tab:student_ablation} is therefore treated as run-to-run
variation, not an independent representational gain. Segmentation supplies
$K$ local distillation units and enables the $K$-sensitivity analysis.

\subsubsection{Effect of Shared Latent-Channel Scaling}

The Reweight module applies an identity-initialized learnable scale to
each latent dimension of the pooled segment features. As discussed in
Section~III, this operation does not recover responses discarded by max
pooling and its coefficients do not correspond directly to the five input
attributes. It instead rescales the contributions of the aggregated
hidden channels before classification and distillation. The scale is shared
across samples and segments and is not a sample-dependent gating mechanism.
As shown in Table~\ref{tab:student_ablation}, at $K=4$, the complete student
with this module has higher mean values (87.04\% accuracy and 87.10\%
macro-F1) than the segment-only variant (86.30\% and 86.13\%, respectively).
This comparison is descriptive and is not presented as a paired statistical
test of the scaling parameters.

\subsubsection{Sensitivity to the Number of Temporal Segments}

We vary $K$ after enabling both pooling and scaling, isolating temporal
granularity from the Reweight module. In Table~\ref{tab:student_ablation},
$K=4$ yields the highest evaluated accuracy and macro-F1. Three segments
may merge distinct phases, whereas five or six create shorter windows
potentially more sensitive to sparse or transient responses. This supports
the default $K=4$, not universal optimality across lengths or datasets.

\subsubsection{Effect of Soft-Align Knowledge Distillation}

Starting from the complete radar student with $K=4$, we first compare two
feature-level alignment strategies. Hard-Align KD divides the teacher
sequence using the same fixed segment boundaries as the student, applies
temporal max pooling within each teacher segment, and matches the
corresponding student and teacher representations. It improves accuracy
from 87.04\% to 87.78\%, but prespecifies which teacher interval should
supervise each student segment.

Soft-Align KD instead allows every student segment to retrieve a weighted
target from the complete teacher sequence. As shown in
Table~\ref{tab:kd_ablation}, it reaches 89.17\% accuracy and 89.18\%
macro-F1, giving higher mean values than both the radar-only student and
fixed segment alignment. This comparison supports adaptive teacher-evidence retrieval
when the two modalities have different representation granularities. The
constrained variant reported in the same table is analyzed separately in
the following subsection.

\begin{table}[h]
\centering
\caption{Distillation Ablation on MM-Fi.}
\label{tab:kd_ablation}
\footnotesize
\setlength{\tabcolsep}{3pt}
\renewcommand{\arraystretch}{1.08}
\begin{tabularx}{\columnwidth}{@{}>{\raggedright\arraybackslash}Xcc@{}}
\hline
Distillation Strategy & Accuracy (\%) & Macro-F1 (\%) \\
\hline
Student & $87.04 \pm 2.97$ & $87.10 \pm 2.75$ \\
Hard-Align KD & $87.78 \pm 4.59$ & $87.79 \pm 4.42$ \\
Soft-Align KD (Ours) & $\mathbf{89.17 \pm 6.71}$ & $\mathbf{89.18 \pm 6.68}$ \\
Constrained Soft-Align KD & $88.52 \pm 6.19$ & $88.40 \pm 6.35$ \\
\hline
\end{tabularx}
\end{table}

\subsubsection{Effect of the Order-and-Diversity Constraint}

\begin{table}[h]
\centering
\caption{Temporal Attention Characteristics on MM-Fi.}
\label{tab:alignment_behavior}
\footnotesize
\setlength{\tabcolsep}{3pt}
\renewcommand{\arraystretch}{1.08}
\begin{tabularx}{\columnwidth}{@{}>{\raggedright\arraybackslash}Xcc@{}}
\hline
Metric & Soft-Align KD & Constrained KD \\
\hline
Duplicate rate (\%) $\downarrow$ & $60.42\pm11.84$ & $\mathbf{7.10\pm3.77}$ \\
Order violation (\%) $\downarrow$ & $50.59\pm1.30$ & $\mathbf{0.37\pm0.35}$ \\
Overlap (0--1) $\downarrow$ & $0.72\pm0.09$ & $\mathbf{0.37\pm0.09}$ \\
Normalized entropy (0--1) & $0.13\pm0.04$ & $0.70\pm0.03$ \\
\hline
\end{tabularx}
\end{table}

\begin{table*}[t]
\centering
\caption{Robustness to Point and Frame Dropout.}
\label{tab:robustness}
\footnotesize
\setlength{\tabcolsep}{4pt}
\renewcommand{\arraystretch}{1.08}
\begin{tabular*}{\textwidth}{@{\extracolsep{\fill}}ccccccc@{}}
\hline
& \multicolumn{2}{c}{Student}
& \multicolumn{2}{c}{Soft-Align KD}
& \multicolumn{2}{c}{Constrained Soft-Align KD} \\
Ratio & Acc. (\%) & Macro-F1 (\%) & Acc. (\%) & Macro-F1 (\%) & Acc. (\%) & Macro-F1 (\%) \\
\hline
\multicolumn{7}{c}{Point dropout} \\
\hline
0.0 & $87.04 \pm 2.97$ & $87.10 \pm 2.75$
    & $\mathbf{89.17 \pm 6.71}$ & $\mathbf{89.18 \pm 6.68}$
    & $88.52 \pm 6.19$ & $88.40 \pm 6.35$ \\
0.2 & $82.22 \pm 3.05$ & $81.99 \pm 3.13$
    & $88.61 \pm 6.02$ & $88.60 \pm 5.82$
    & $\mathbf{89.17 \pm 6.62}$ & $\mathbf{88.93 \pm 6.90}$ \\
0.4 & $79.26 \pm 5.78$ & $79.03 \pm 5.15$
    & $87.31 \pm 6.35$ & $87.33 \pm 6.14$
    & $\mathbf{88.06 \pm 5.57}$ & $\mathbf{87.78 \pm 5.68}$ \\
0.6 & $70.19 \pm 8.43$ & $69.44 \pm 8.44$
    & $\mathbf{80.46 \pm 6.80}$ & $\mathbf{80.33 \pm 6.74}$
    & $80.09 \pm 5.85$ & $79.57 \pm 5.92$ \\
\hline
\multicolumn{7}{c}{Frame dropout} \\
\hline
0.0 & $87.04 \pm 2.97$ & $87.10 \pm 2.75$
    & $\mathbf{89.17 \pm 6.71}$ & $\mathbf{89.18 \pm 6.68}$
    & $88.52 \pm 6.19$ & $88.40 \pm 6.35$ \\
0.1 & $83.89 \pm 5.09$ & $84.14 \pm 4.48$
    & $\mathbf{85.74 \pm 3.38}$ & $\mathbf{85.73 \pm 2.94}$
    & $85.46 \pm 7.21$ & $85.23 \pm 7.38$ \\
0.2 & $80.74 \pm 3.88$ & $80.34 \pm 3.83$
    & $\mathbf{85.19 \pm 4.46}$ & $\mathbf{85.19 \pm 4.28}$
    & $84.35 \pm 6.96$ & $84.22 \pm 7.05$ \\
0.3 & $78.33 \pm 7.27$ & $77.80 \pm 7.68$
    & $84.91 \pm 4.20$ & $85.05 \pm 3.75$
    & $\mathbf{85.19 \pm 6.44}$ & $\mathbf{85.06 \pm 6.59}$ \\
\hline
\end{tabular*}
\end{table*}

Independent attention normalization permits shared evidence and
non-monotonic retrieval, which may benefit classification rather than
indicate alignment errors. An auxiliary constrained variant tests whether
these patterns can be explicitly controlled.

The following per-sample losses are averaged over the mini-batch. Let
$r_t=(t-1)/(T-1)\in[0,1]$ denote normalized teacher-frame position,
with $T>1$ teacher states and $K\geq2$ student segments. Since
$\alpha_{k,t}\geq0$ and $\sum_{t=1}^{T}\alpha_{k,t}=1$, the weights
form a distribution over teacher time. Its expected position
$\rho_k\in[0,1]$ and adjacent-center gap $g_k$ are

\begin{equation}
\rho_k=\sum_{t=1}^{T}\alpha_{k,t}r_t,
\qquad
g_k=\rho_{k+1}-\rho_k .
\end{equation}

Thus, $g_k>0$ places segment $k+1$'s center later than segment $k$'s;
$g_k\leq0$ indicates an adjacent order violation.

An order-and-separation penalty encourages adjacent centers to increase by
at least a margin $\delta$:

\begin{equation}
\mathcal{L}_{\mathrm{ord}}
=
\frac{1}{K-1}
\sum_{k=1}^{K-1}
\left(
\frac{[\delta-g_k]_+}{\delta}
\right)^2 .
\end{equation}

Here, $[x]_+=\max(x,0)$ and $\delta>0$ is the minimum gap: the penalty
is zero for $g_k\geq\delta$ and increases for closer or reversed centers.

To discourage overlapping teacher-time distributions, define
$\alpha_k=[\alpha_{k,1},\ldots,\alpha_{k,T}]^{\mathsf{T}}\in\mathbb{R}^{T}$
as the complete temporal attention vector of segment $k$; $\alpha_l$
denotes the corresponding vector of another segment $l\neq k$. We
minimize their mean off-diagonal cosine similarity:

\begin{equation}
\begin{aligned}
\mathcal{L}_{\mathrm{div}}
&=
\frac{1}{K(K-1)}
\sum_{k\neq l}
\frac{\alpha_k^{\mathsf{T}}\alpha_l}
{\|\alpha_k\|_2\|\alpha_l\|_2},\\
\mathcal{L}_{\mathrm{con}}
&=
\mathcal{L}_{\mathrm{ord}}
+\beta\mathcal{L}_{\mathrm{div}} .
\end{aligned}
\end{equation}

The dot product $\alpha_k^{\mathsf{T}}\alpha_l$ is normalized by both
$\ell_2$ magnitudes. For nonnegative attention, larger cosine similarity
indicates stronger temporal overlap. $\mathcal{L}_{\mathrm{div}}$
averages it over all ordered pairs $k\neq l$; $\beta$ weights diversity
relative to the order penalty.

This regularizer encourages ordered and less-overlapping use of teacher
evidence. It does not impose hard one-to-one correspondence, anchor the
first and last attention centers to the sequence boundaries, or guarantee
complete coverage of every teacher frame.

The constrained model is trained from scratch using
$\mathcal{L}_{CE}+0.5\mathcal{L}_{KD}+\gamma(e)\mathcal{L}_{\mathrm{con}}$. Here, $e$ is the training epoch and $\gamma(e)$ is the epoch-dependent weight of the auxiliary constraint; $\mathcal{L}_{CE}$ and
$\mathcal{L}_{KD}$ retain the definitions used by the main Soft-Align
model. We set the minimum center gap to one half of the nominal inter-segment spacing, i.e., $\delta=0.5/K$. Thus, for $K=4$,
$\delta=0.125$ on the normalized teacher timeline. We set $\beta=0.1$. To avoid disturbing early representation learning, $\gamma(e)$ is zero for the first 10 epochs, increases linearly to 0.05 over epochs 11--20, and remains at 0.05 thereafter. The gradient norm is clipped at 5.0. Thus, this experiment uses a single from-scratch training schedule with loss warm-up rather than fine-tuning an unconstrained checkpoint.

As shown in Table~\ref{tab:kd_ablation}, the constrained variant achieves 88.52\% accuracy and 88.40\% macro-F1. It remains above the radar-only student and Hard-Align KD, but is 0.65 percentage points below the unconstrained Soft-Align KD in accuracy. Table~\ref{tab:alignment_behavior} confirms that the auxiliary objective substantially changes the alignment behavior.

Duplicate rate is the fraction of segment pairs peaking at the same
teacher frame; order-violation rate is the fraction of adjacent centers
with $g_k\leq0$; overlap is their mean pairwise attention cosine
similarity; normalized entropy averages
$-\sum_t\alpha_{k,t}\log\alpha_{k,t}/\log T$ over segments, where
$\log$ denotes the natural logarithm. Larger values indicate broader
attention. The constraint reduces duplication, order
violations, and overlap while broadening attention. Its small accuracy
decrease shows that these interpretability-related properties do not
coincide with the classification objective. We retain unconstrained
Soft-Align KD as the main method and the constraint as a mechanism ablation.

\subsection{Robustness Experiment}

To evaluate robustness to degraded radar observations, we apply two
types of random corruption only at inference time: point dropout with
ratios $\{0.0,0.2,0.4,0.6\}$ and frame dropout with ratios
$\{0.0,0.1,0.2,0.3\}$. The model parameters remain fixed. We compare the
radar-only student, the main unconstrained Soft-Align KD model, and the
order-and-diversity constrained variant introduced in the preceding
ablation.

\begin{figure}[!t]
\centering
\includegraphics[width=\linewidth]{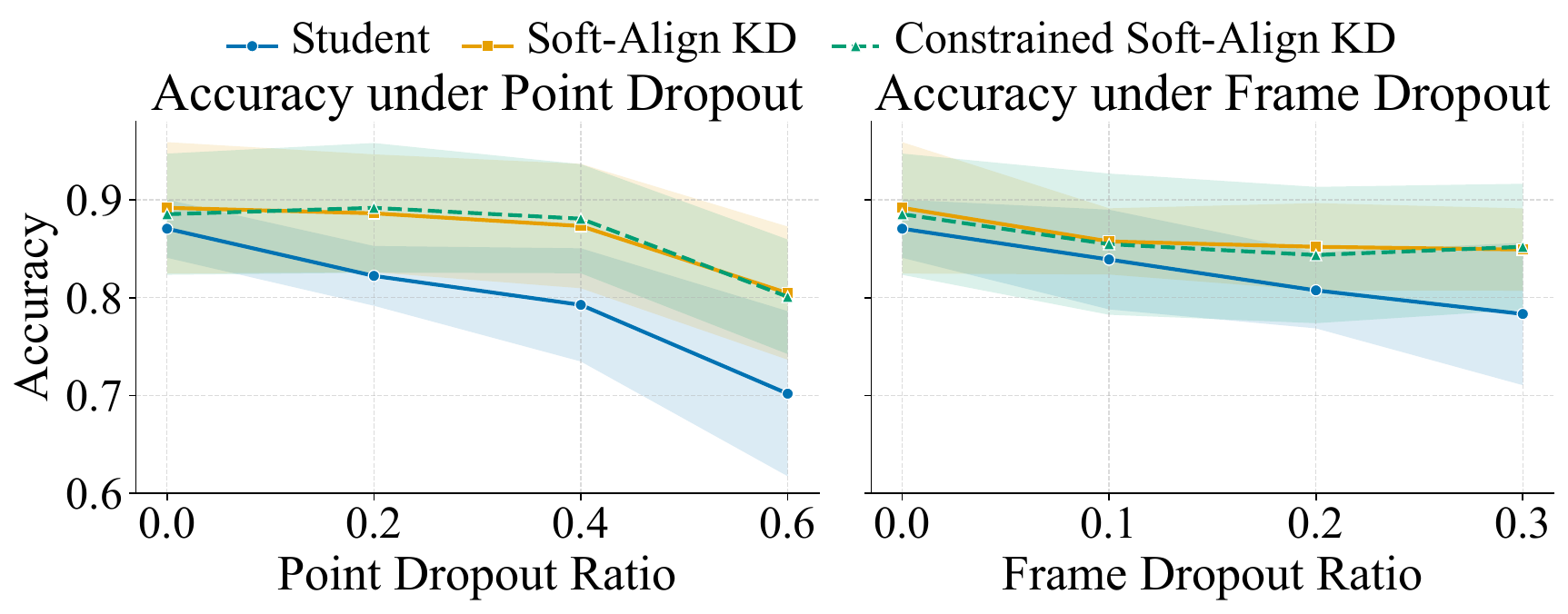}
\caption{Accuracy of the student, Soft-Align KD, and constrained
Soft-Align KD under random point dropout (left) and frame dropout
(right). Markers and shaded regions represent the five-run mean and one
standard deviation, respectively.}
\label{fig:robustness_accuracy}
\end{figure}

\begin{figure}[!t]
\centering
\includegraphics[width=\linewidth]{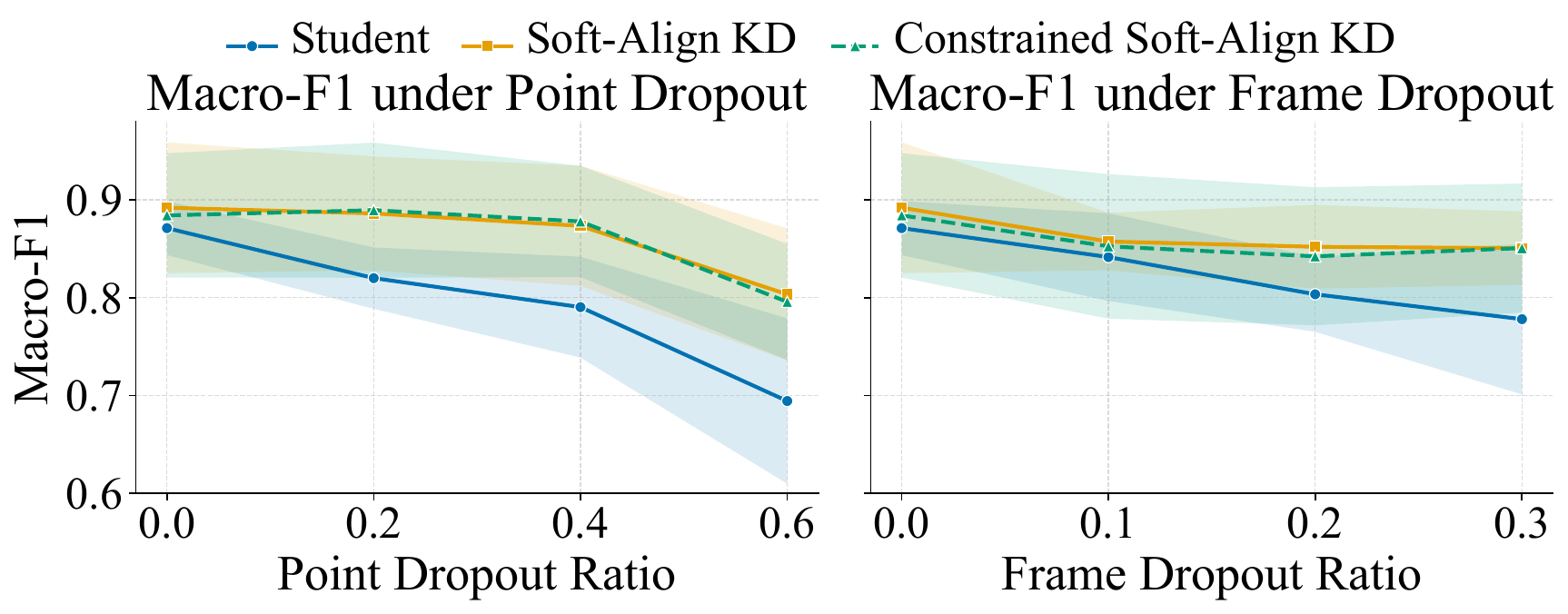}
\caption{Macro-F1 of the student, Soft-Align KD, and constrained
Soft-Align KD under random point dropout (left) and frame dropout
(right). Markers and shaded regions represent the five-run mean and one
standard deviation, respectively.}
\label{fig:robustness_f1}
\end{figure}

Table~\ref{tab:robustness} reports the five development-fold results. Both distilled
models obtain higher mean scores than the radar-only student under most
corruption levels. In particular, at 0.6 point dropout, unconstrained
Soft-Align KD raises mean accuracy from 70.19\% to 80.46\% and mean macro-F1
from 69.44\% to 80.33\%. At 0.3 frame dropout, the corresponding
mean changes are from 78.33\% to 84.91\% in accuracy and from 77.80\%
to 85.05\% in macro-F1.

Figs.~\ref{fig:robustness_accuracy} and
\ref{fig:robustness_f1} visualize the same trends. The constrained
variant gives the highest means at point-dropout ratios 0.2 and 0.4 and
is marginally higher at frame-dropout ratio 0.3. In contrast, the
unconstrained model is better on clean data, at severe point dropout
(0.6), and at frame-dropout ratios 0.1 and 0.2. Moreover, the
run-to-run standard-deviation ranges substantially overlap. The
constraint therefore does not provide a consistent robustness advantage.
Together with its lower clean-set accuracy, this result supports retaining
unconstrained Soft-Align KD as the main classification method and treating
the constrained formulation as a mechanism-oriented ablation.

\subsection{Cross-Dataset Adaptation Experiment}
\label{sec:generalization}

With the same student architecture, we compare training from scratch,
MM-Fi radar-only pretraining, and unconstrained Soft-Align KD pretraining
on mmO-HAR. Pretrained strategies transfer compatible parameters and
initialize a new target classifier, following the subject-disjoint test
protocol in Section~\ref{sec:mmO-HAR}.

The mmO-HAR setup differs from MM-Fi in sensor type (TI IWR6843ISK versus
TI IWR6843AOP), environment (an office versus multiple indoor scenes),
activity categories (6 versus 27), and sequence length (60 versus 297
frames). In Table~\ref{tab:mmO_HAR_results}, Soft-Align KD pretraining
achieves 87.75\% accuracy and 86.74\% macro-F1: gains of 7.42 and 6.26
percentage points over training from scratch, and 5.67 and 4.94 over
radar-only pretraining. This supports pose-supervised radar initialization
for the independently collected target, not universal cross-dataset
generalization.

\begin{table}[!htbp]
\centering
\caption{Transfer Learning Results on mmO-HAR.}
\label{tab:mmO_HAR_results}
\footnotesize
\setlength{\tabcolsep}{3pt}
\renewcommand{\arraystretch}{1.08}
\begin{tabularx}{\columnwidth}{@{}>{\raggedright\arraybackslash}Xcc@{}}
\hline
Initialization strategy & Accuracy (\%) & Macro-F1 (\%) \\
\hline
Student (scratch) & $80.33 \pm 4.19$ & $80.48 \pm 4.02$ \\
+ MM-Fi radar pretraining & $82.08 \pm 5.16$ & $81.80 \pm 5.43$ \\
+ Soft-Align KD pretraining & $\mathbf{87.75 \pm 6.14}$ & $\mathbf{86.74 \pm 6.94}$ \\
\hline
\end{tabularx}
\end{table}

\FloatBarrier
\section{Conclusion}

We presented cross-modal distillation from a ResNet-18--BiLSTM keypoint
teacher to a radar student combining point-set encoding, bidirectional
temporal modeling, segment pooling, and shared latent-channel scaling.
Soft-Align KD retrieves a contextualized teacher target for each student
segment rather than imposing hard interval correspondence. This is
student-conditioned, potentially non-monotonic evidence retrieval, not
chronological registration. Removing the teacher and alignment projection
leaves radar-only inference with 5.568~M parameters and 16.020~GMACs per
sequence, identical to the independently trained student.

On MM-Fi, Soft-Align KD raises accuracy/macro-F1 from 87.04\%/87.10\%
to 89.17\%/89.18\%, the highest evaluated means; WKD-L reaches
88.15\%/88.31\%. The five development-fold runs share a fixed test set,
so these are numerical mean improvements, not uniform fold-wise or
statistically significant superiority. Distillation also improves mean
scores over the student under most point- and frame-dropout settings.
On the subject-disjoint mmO-HAR test set, pretraining improves accuracy
from 80.33\% to 87.75\%, supporting adaptation of pose-supervised radar
representations to an independently collected target.

The alignment audit confirms that segments may reuse teacher context.
An order-and-diversity constraint reduces duplication, order violations,
and overlap but lowers clean-set performance and provides no consistent
robustness gain, revealing a trade-off between classification-oriented
aggregation and ordered, interpretable correspondence. A100 profiling
does not establish embedded-device latency or energy consumption. Future
work includes adaptive temporal granularity, weak structure-aware
objectives, broader cross-domain evaluation, deterministic repeated
trials, and on-device evaluation of lighter radar models.

\bibliographystyle{IEEEtran}
\bibliography{reference}
\end{document}